# Is La$_3$Ni$_2$O$_{6.5}$ a Bulk Superconducting Nickelate?

Ran Gao[1#], Lun Jin[1#], Shuyuan Huyan[2,3#], Danrui Ni[1*], Haozhe Wang[4], Xianghan Xu[1], Sergey L. Bud'ko[2,3], Paul Canfield[2,3], Weiwei Xie[4*] and Robert J. Cava[1*]

[1]Department of Chemistry, Princeton University, Princeton, New Jersey 08544, USA

[2]Ames National Laboratory, Iowa State University, Ames, IA 50011, USA

[3]Department of Physics and Astronomy, Iowa State University, Ames, IA 50011, USA

[4]Department of Chemistry, Michigan State University, East Lansing, Michigan 48824, USA

* E-mails of corresponding authors: xieweiwe@msu.edu; danruin@princeton.edu; rcava@princeton.edu

\# L.J., R.G. and S.H. contributed equally.

*Abstract*

Superconducting states onsetting at moderately high temperatures have been observed in epitaxially-stabilized $RE$NiO$_2$-based thin films. However, recently it has also been reported that superconductivity at high temperatures is observed in bulk La$_3$Ni$_2$O$_{7-\delta}$ at high pressure, opening further possibilities for study. Here we report the reduction profile of La$_3$Ni$_2$O$_7$ in a stream of 5% H$_2$/Ar gas and the isolation of the metastable intermediate phase La$_3$Ni$_2$O$_{6.45}$, which is based on Ni$^{2+}$. Although this reduced phase does not superconduct at ambient or high pressures, it offers insights into the Ni-327 system and encourages the future study of nickelates as a function of oxygen content.

**Introduction**

Since the discovery of superconducting cuprates late in the last century [1], nickelates have been predicted to potentially host this exotic state of matter [2], primarily because Ni, a magnetic element, is one of the closest neighbors of Cu on the periodic table. Despite much experimental effort on nickelates [e.g. 3,4], superconductivity in epitaxially-stabilized doped $RE$NiO$_2$ thin films (topochemically reduced from the parent 113 perovskites) has only recently been discovered [5–8]. This superconductivity has not to date been seen in bulk materials. Very recently, however, superconductivity near 80 K has been reported for La$_3$Ni$_2$O$_{7-\delta}$ in bulk form at high pressures. [9–12] This advancement, if true, not only extends superconducting nickelates to bulk materials with a much higher critical temperature, but also paves the way towards more possible candidate materials that should be examined. While some may argue that we have thus entered the "nickelate superconductor age", there are still many challenges ahead before that can be taken as a certainty.

In the present study, we characterize the reduction of the bilayer nickelate La$_3$Ni$_2$O$_7$, using thermogravimetric analysis. Its reduction in forming gas (in our case 5% H$_2$ in Ar) while maintaining its basic crystal structure [13–15] is highly unusual for a transition metal oxide, which is due to the stability of Ni$^{2+}$ in oxides, yielding a better than average possibility for changing its carrier concentration. In addition to the equilibrium behavior, we report an unusual kinetic effect on the phase assemblage observed on reduction. Finally, we have successfully isolated the metastable intermediate phase La$_3$Ni$_2$O$_{6.45}$, a composition that is a member of the La$_3$Ni$_2$O$_{7-\delta}$ family and have investigated its structural and physical properties. By removing about 0.5 oxygen per formula unit from the parent compound, we have introduced an intrinsic magnetic transition instead of superconductivity. We have looked for signs of superconductivity in this reduced phase at both ambient and high pressures but have not seen any. Even though the superconductivity is absent in our material, our results indicate that the investigation of difficult-to-reduce oxides should be of future interest.

**Experimental**

Polycrystalline samples of La$_3$Ni$_2$O$_7$ were synthesized by a high-temperature ceramic method using La$_2$O$_3$ (Alfa Aesar, 99.99%) and NiO (Sigma-Aldrich, ~325 mesh, 99% with 1% unspecified impurity) as the starting materials. La$_2$O$_3$ was pre-dried overnight in a furnace at 900 °C. Stoichiometric amounts (in metals) of the starting materials were weighed accurately and ground together thoroughly before being transferred into an alumina crucible. The mixed starting materials were reacted in a furnace at 1100 °C for 5 days in air with intermittent grinding. Both heating and cooling were set at a rate of 3 °C per minute. Reduced samples with a series of formulas of the type $La_3Ni_2O_{7-x}$ were attained by the thermogravimetric analysis (TGA) technique with a TA Instruments TGA 5500. The parent $La_3Ni_2O_7$ phase was reduced in a forming gas of 5% $H_2$ in Ar at different temperatures ranging from 300 °C to 800 °C and held isothermally for various periods of time. A ramp rate of 10 °C per minute was used during heating, and fast cooling to ambient temperature was used after the isothermal segment.

Powder X-ray diffraction (PXRD) measurements were performed by using a Bruker D8 FOCUS diffractometer (Cu Kα radiation) at ambient temperature. PXRD data were collected after each intermittent grinding and each TGA reduction to monitor the reaction process. Once the pure phase was attained, the PXRD data were collected at room temperature over a 2θ range from 5 ° to 110 °, with much better statistical significance, for the following Le Bail refinements conducted by using TOPAS software.

The physical properties of all polycrystalline samples were analyzed by a Quantum Design Dynacool Physical Property Measurement System (PPMS). Magnetic measurements were performed by the PPMS, which was equipped with a vibrating sample magnetometer (VSM). Magnetic susceptibility (χ) is defined as *M/H*, where *M* is magnetization and *H* is the applied magnetic field intensity. Heat capacity was measured by a standard relaxation method over a temperature range from 2 K to 70 K under 0 T field.

High pressure electrical resistivity measurements using the Van der Pauw method were performed in a commercial Diamond Anvil cell (DAC) [16] that fits the Quantum Design Physical Property Measurement System (PPMS). Standard cut-type Ia diamonds with a culet size of 400 μm were utilized as the anvils, while a 250 μm thick stainless-steel aperture disc served as the gasket. To ensure electrical insulation, the aperture was filled and compressed with a mixture of cubic boron nitride (cBN) and epoxy, and the rest of gasket's top surface was covered by STYCAST. Subsequently, a central hole with a diameter of 150 μm was drilled through the cBN layer to create a sample chamber. This chamber was initially loaded with fine powder of NaCl, which acted as the pressure-transmitting medium (PTM). The NaCl powder was then compressed to ~ 1 to 2 GPa until the hole became totally transparent. After that, an additional, approximately 80 μm half-drilled hole was created within the NaCl layer. The $La_3Ni_2O_{7-\delta}$ powder sample was carefully introduced and pressed (2.1GPa) into the half-drilled hole, with a small ruby sphere placed at the bottom of the sample as the manometer. Pressure was determined by the $R_1$ line position of the ruby fluorescent spectra [17]. Platinum foil electrodes were employed to establish electrical connections with the sample.

**Results and Discussion**

Reduced phases of $La_3Ni_2O_7$ were prepared by using different TGA sequences, with the weight % variations plotted against both temperature and time in **Figure 1a&b**. In general, the as-made $La_3Ni_2O_7$ phase was reduced in a forming gas of 5% $H_2$ in Ar during the isothermal segment that lasts for 3-10 hours, followed by a fast-cooling process. As shown in **Figure 1**, the as-made $La_3Ni_2O_7$ phase started to lose mass continuously at 300 °C and didn't reach a final, stable oxygen content even after a 10-hour isothermal reduction period. Therefore, it was necessary to increase the reduction temperature. A long-time stable weight occurred when the dwelling temperature was set to 350 °C, 400 °C, and 450 °C, respectively. Although a dwelling period of 10 h was required for the 350 °C sequence to reach a relatively constant mass, the samples reduced at 400 °C and 450 °C reached a stable mass within a 4–6 hour isothermal reduction period. When the reduction temperature was elevated to 500 °C, the $La_3Ni_2O_7$ sample reached the same stable mass first, and after staying at this plateau for approximately 2 hours, a further mass loss began. Thus at 500 C, the stable mass material is attainable but there is a clear kinetic effect at play. The weight loss curve didn't reach a new stable value after the 6-hour isothermal reduction at 500 °C. For full reduction and decomposition to $La_2O_3$ plus elemental Ni, the $La_3Ni_2O_7$ sample was reduced at 800 °C, and the weight change curve reached a new platform with a 3-hour isothermal step, corresponding to an approximately 6% loss of the total weight. Although the "intermediate mass platform" can still be vaguely spotted in the 800 °C curve, it is obviously shorter compared to that present for the lower temperature reductions. From the weight % versus time plot (**Figure 1b**), we can observe that higher temperatures speed up the reduction process. In addition to the reduction temperature, the annealing time also contributes to the reduction process of $La_3Ni_2O_7$, which is most obvious from in the 500 °C TGA run, as the "intermediate platform mass" only lasted for 2 hours, while further annealing at this temperature gradually destroys this phase, making it metastable. By using PXRD to examine the sample residue after each TGA run, we found that the structural lattice of $La_3Ni_2O_7$ had collapsed and reduced to the non-topochemical decomposition products $La_2O_3$ and Ni at 800 °C, hence the formula of the as-made sample, $La_3Ni_2O_{7.01}$, can be calculated precisely based on the weight loss from the 800 °C TGA curve. Therefore, the oxygen stoichiometry of the intermediate topochemically reduced phase is determined as $La_3Ni_2O_{6.45}$ based on the weight % loss at the first plateau, which can be clearly observed in the TGA curves obtained at 400 °C and 450 °C (**Figure 1a**). This composition we take as being within error of $La_3Ni_2O_{6.5}$, which may be metastable for this material. At this composition the Ni would all be present in the 2+ state, a highly stable $d^8$ configuration for the ionized nickel [18,19].

PXRD patterns of the post-reduction powder samples were collected to analyze the structural change, and Le Bail fittings were conducted on the as-made and reduced phases to determine the symmetry and size of the unit cells. The as-made parent $La_3Ni_2O_{7.01}$ phase can be indexed by a face-centered orthorhombic unit cell (space group *Fmmm*) with lattice parameters $a$ = 5.3924(2) Å, $b$ = 5.4474(2) Å, and $c$ = 20.532(1) Å (**Figure 2a**), in good agreement with previous reports [13]. The body-centered tetragonal unit cell (space group *I4/mmm*) that has been widely adopted for reduced phases of $La_3Ni_2O_{7-\delta}$ in the literature [14,15], can be used to index our metastable reduced phase $La_3Ni_2O_{6.45}$ as well, with lattice parameters $a$ = 3.8734(2) Å and $c$ = 20.075(1) Å (**Figure 2b**). This transition in symmetry from orthorhombic to tetragonal, caused by the oxygen deficiency, is consistent with previous reports [13,14]. It is noted that after the $\sqrt{2}$ expansion of the tetragonal unit cell of the reduced phase, its parameter $a$ (5.48 Å) is only marginally enlarged compared to the parameters $a$ and $b$ of the orthorhombic unit cell of the parent compound, while parameter $c$ shrinks approximately 0.5 Å upon the reduction. Based on the published structural refinements [13,14], the oxygen vacancies in the reduced phases are not randomly distributed. In general, they prefer to sit in the bridging apical oxygen site that connects the two Ni centers along the $c$-axis, in the La-O planes (The 4$a$ site in the *Fmmm* unit cell and the 2$a$ site in the *I4/mmm* unit cell). This can also explain the larger change in lattice parameter $c$ on reduction.

Using the results of the PXRD data collected from the post-TGA samples reduced at different temperatures, more detailed insight into the reduction behavior of $La_3Ni_2O_7$ can be revealed. As shown in Figure 2(c), the reduction started at 300 °C in the 5% $H_2$/Ar forming gas, though La$_3$Ni$_2$O$_7$ didn't fully convert to the tetragonal reduced phase after a 10-hour isothermal process. The incomplete structural transition can be diagnosed by the partial merging of peaks (such as the doublets at 27, 33, 43, and 58 degrees) and the asymmetric peak shape (near 47 degrees) when comparing the black and red patterns in **Figure 2c**, which also agrees well with the TGA results. The highly similar PXRD patterns of the post-TGA samples between 350 °C and 450 °C align well with the constantly observed metastable reduced La$_3$Ni$_2$O$_{7-\delta}$ phase in those TGA runs, indicating that the samples reduced at those temperatures have the same composition, La$_3$Ni$_2$O$_{6.45}$. In the 500 °C pattern, the peaks representing La$_2$O$_3$ and Ni start to show up, while the crystallinity of the target La$_3$Ni$_2$O$_{6.45}$ phase is highly compromised. This observation is mutually supported by the second weight % drop after the 2-hour plateau in the 500 °C TGA curve, confirming that the structural lattice disintegrates upon prolonged heating at this temperature. The peaks of the La$_3$Ni$_2$O$_{6.45}$ phase completely disappear in the 800 °C pattern; the pattern is now composed of only La$_2$O$_3$ and Ni peaks, indicating a thorough decomposition of the ternary compound. The continuously evolving peaks in the PXRD patterns from the as-made sample to the 450 °C reduced one reveal that the structural transition occurs through a series of compositions with a short-range distribution of different oxygen contents, which is clearly distinguishable from the final non-topochemical decomposition of the structural lattice (patterns 500 °C to 800 °C).

Temperature-dependent magnetization data ($M$) were collected from the reduced phase La$_3$Ni$_2$O$_{6.45}$, synthesized at 400 °C, using both zero-field-cooled (ZFC) and field-cooled (FC) approaches using an applied field of $H$ = 1000 Oe. To avoid any potential ambiguities, the collected magnetization data were all scaled to per-mole-Ni (based on the formula La$_{1.5}$NiO$_{3.225}$). The resulting magnetic susceptibility ($M/H$) data are plotted against temperature in **Figure 3a**. The ZFC and FC curves diverge from each other in the high-temperature regime. The ZFC curve distinctively shows two transitions - at temperatures of about 10 K and 50 K, with the latter being much broader than the former. The field-dependent data were collected from the reduced phase at both 5 K and 300 K over the field range -9 T $\leq H \leq$ 9 T. The 300 K magnetization isotherm exhibits a "kink" instead of being a straight line passing through the origin, indicating that a trace amount (~ 0.43% per mole of La$_{1.5}$NiO$_{3.225}$, calculated by comparing the 300 K isotherm step heights in **Figure 3b** and **S2**) of ferromagnetic impurity with ~0.43% molar ratio when compared to that of elemental Ni exists in the reduced sample La$_3$Ni$_2$O$_{6.45}$, which is a known situation for topochemically reduced Ni-containing oxides [20–22]. The subtly opened-up hysteresis loop in the 5 K isotherm further confirms the existence of ferromagnetic impurity in the system, which is equivalent to 0.43% by mole of elemental Ni (**Figure 3b**). Therefore, it is necessary to employ a ferro-subtraction technique, which can unveil the magnetic behavior of the bulk material by applying an external field greater than 2 T to saturate the magnetization of the ferromagnetic impurity. Thus in 5 K increments between 5 K and 300 K, field-dependent magnetization data were collected over the field range $3\,T \leq H \leq 5\,T$ and the resultant magnetizations ($M$) plotted as a function of the applied field ($H$). For each temperature, the slope of the plot obtained (i.e., $\Delta M/\Delta H$) by linear-fitting the data over the field range $3\,T \leq H \leq 5\,T$ can be considered as the paramagnetic susceptibility of the bulk material, since a saturation stage of ferromagnetic impurity starts to occur at field much smaller than 3 T (**Figure S1-2**). Thus, a plot can be made by combining the data that was taken over the whole temperature range, which shows the temperature dependence of the bulk material's susceptibility. In **Figure 3c**, the kink around 10 K may either be from the ferromagnetic impurity or a transition of some kind in the bulk material. The ferro-subtracted magnetic susceptibility curve shows one major phase transformation starting at $T \approx 50$ K, which suggests that the bulk material has a relatively broad transition. A modified Curie-Weiss law ($\chi = C/(T - \theta) + \chi_0$) was used to fit the high-temperature region (100 - 250 K) of the inverse magnetic susceptibility data, to derive the Curie constant and the Curie-Weiss temperature of the bulk material (**Figure 3d**). The fitted Curie-Weiss temperature $\theta$ is 7.5(8) K, indicating the intrinsic weakly ferromagnetic average coupling in bulk $La_3Ni_2O_{6.45}$. The calculated value of Curie constant $C$ is 0.0705 $cm^3\,K\,mol_{Ni}^{-1}$ and yields an effective moment of 0.75 $\mu_B$ per Ni. The average oxidation state of Ni in the reduced phase is +1.95 according to the nominal composition, suggesting a combination of $Ni^{1+}$ and $Ni^{2+}$ in a localized picture. The theoretical effective moment per Ni is 1.73 $\mu_B$ for $Ni^{1+}$ ($S = 1/2$) and 2.83 $\mu_B$ for $Ni^{2+}$ ($S = 1$) based on the spin-only values for localized spins, which are significantly higher than the experimental value. The small magnitude of the experimental effective moment and the fact that the Curie-Weiss temperature is apparently lower than the magnetic ordering temperature, accompanied by the relatively broad shape of the observed transition in the ferro-subtracted magnetic susceptibility curve, indicate that future work on the magnetism of this material would be of interest.

Since we can still observe a broad hump in the magnetic susceptibility data after eliminating the contribution of a trace amount (~ 0.43% per mole of $La_{1.5}NiO_{3.225}$) of ferromagnetic impurity, heat capacity data were then collected from a cold-pressed pellet of $La_3Ni_2O_{6.45}$, to further investigate the nature of this anomaly. The collected heat capacity data were scaled to per-mole-Ni (based on the formula $La_{1.5}NiO_{3.225}$). The reduced phase $La_3Ni_2O_{6.45}$ shows semiconducting behavior, consistent with reports in literature [23], and thus there are few conduction electrons present that would yield a significant $C_{electron}$. The total heat capacity data ($C_{total}$) are plotted against temperature T in **Figure 4a**. There is no clearly defined peak in the heat capacity data and therefore the excess heat capacity collected under zero field in the temperature range of 50 K to 70 K (above the magnetic transition), is estimated by fitting to a modified two-component Debye equation $C_{phonon} = 9R \sum_{n=1}^{2} C_n \left(\frac{T}{\Theta_{Dn}}\right)^3 \int_0^{\Theta_{Dn}/T} \frac{x^4 e^x}{(e^x-1)^2} dx$, in where $C_1 = 3.225$, $\Theta_{D1} = 734(4)$ K, $C_2 = 2.5$, and $\Theta_{D2} = 258(1)$ K, to estimate the phonon contribution, $C_{phonon}$. The $C_1$ and $C_2$ values are in good agreement with the 2.5 heavy atoms (La and Ni) and 3.225 light atoms (O) in the formula $La_{1.5}NiO_{3.225}$. Note that the same two-component Debye equation has been applied to fit the high-temperature heat capacity of other oxide solids with coexisting light and heavy atoms [24,25]. Therefore, after subtracting the phonon contribution $C_{phonon}$ from $C_{total}$, the excess heat capacity contribution $C_{XS}$ can be estimated. The resulting $C_{XS}/T$ (in orange) is plotted against temperature T in **Figure 4b**. The $C_{XS}/T$ curve shows a rise below 10 K, and a broad hump that starts at a similar temperature (50 K) to what we observed in the ferro-subtracted magnetic susceptibility data. Since magnetism is expected for a $Ni^{2+}$-based system, we can reasonably attribute the heat capacity excess to a magnetic transition and as such the magnetic entropy change $\Delta S_{xs}$ (in purple) can be calculated by taking the integral of $C_{XS}/T$ over the measured temperature range, yielding a saturation value of ~ 0.75 $J/mol_{Ni}/K$, which is small even when compared to the Ising spin prediction $R \ln(2) = 5.76$ J/mol/K. Whereas the heat capacity data does suggest that the anomaly seen in the temperature dependent magnetization data is magnetic in origin, more work will be needed to determine nature of the apparent magnetic ordering at low temperature if present.

The reduced phase $La_3Ni_2O_{6.45}$ prepared in the present work is a semiconductor at ambient pressure, consistent with previous descriptions [23,26]. However, given the fact that $La_3Ni_2O_{6.93}$ was recently reported to show signatures of superconductivity under high applied pressure [12], the temperature-dependent resistivity of $La_3Ni_2O_{6.45}$ under pressure is of interest as well. Unfortunately, no signs of superconductivity have been detected for our $La_3Ni_2O_{6.45}$ under applied pressures of up to 41 GPa. The material remains a semiconductor when pressurized, which we deduce from the fact that the material shows an increase in resistance on a decrease of the temperature under all pressures studied (**Figure 5a**). The resistance at ambient temperature is found to drop significantly under applied pressure, with the largest changes occurring below 9 GPa (**Figure 5b**). Also, the semiconducting gap (**Figure 5c**) obtained from the fits of the high temperature part of resistance curves (**Figure S2**) decreases under pressure by factor of ~3., suggesting that applied pressure brings $La_3Ni_2O_{6.45}$ significantly closer to a metallic state than it is at ambient pressure.

## Conclusion

In the present work, the reduced phase $La_3Ni_2O_{6.45}$ was synthesized from the parent $La_3Ni_2O_7$ phase by heating in the TGA under different reduction conditions. The reduction process was studied through the combination of PXRD and TGA data, offering insights into the structural changes between $La_3Ni_2O_7$ and $La_3Ni_2O_{6.45}$. The intrinsic magnetic properties of bulk $La_3Ni_2O_{6.45}$ were unveiled by saturating the inevitably present ferromagnetic impurity under a high external applied field, using a so-called ferro-subtraction technique. The broad hump observed in the ferro-subtracted magnetic susceptibility curve starting at temperature around 50 K, accompanied by the broad anomaly at a similar temperature in the $C_{XS}/T$ curve, confirms the intrinsic behavior of the reduced material. No sign of superconductivity was found in our reduced phase $La_3Ni_2O_{6.45}$, although for superconductors it has been demonstrated that signs of superconductivity can be observed even when the composition of the superconductor is not the composition of the sample, due to the inevitable composition variations present in real materials. Instead, we introduce an intrinsic transition to the otherwise weakly paramagnetic $La_3Ni_2O_7$ and shed light on the reduction profile of the parent oxide compound, which is currently in the spotlight due to recent reports of it displaying high pressure superconductivity. Our work shows that the scientific community should consider whether both the amount and arrangement of oxygen vacancies may be crucial for the nickelates to hold superconductivity, Thus, our study encourages the more detailed study of formulas with different oxygen contents both in nickelates and other oxides that are challenging to stably obtain.

## Acknowledgements

This research was supported by the US Department of Energy, grants DE-FG02-98ER45706 (Princeton University) and DE-SC0023648 (Michigan State University). The work in Ames was supported by the U.S. Department of Energy, Office of Science, Basic Energy Sciences, Materials Sciences and Engineering Division. Ames National Laboratory is operated for the U.S. Department of Energy by Iowa State University under Contract No. DE- AC02-07CH11358.

## Supporting Information

Supporting Information is available from the publisher or from the authors upon valid request.

# References

[1] J. G. Bednorz and K. A. Müller, *Possible high $T_c$ Superconductivity in the Ba−La−Cu−O System*, Z. Für Phys. B Condens. Matter **64**, 189 (1986).

[2] V. I. Anisimov, D. Bukhvalov, and T. M. Rice, *Electronic Structure of Possible Nickelate Analogs to the Cuprates*, Phys. Rev. B **59**, 7901 (1999).

[3] J. Zhang, A. S. Botana, J. W. Freeland, D. Phelan, H. Zheng, V. Pardo, M. R. Norman, and J. F. Mitchell, *Large Orbital Polarization in a Metallic Square-Planar Nickelate*, Nat. Phys. **13**, 9 (2017).

[4] J. Zhang, Y.-S. Chen, D. Phelan, H. Zheng, M. R. Norman, and J. F. Mitchell, *Stacked Charge Stripes in the Quasi-2D Trilayer Nickelate $La_4Ni_3O_8$*, Proc. Natl. Acad. Sci. **113**, 8945 (2016).

[5] D. Li, K. Lee, B. Y. Wang, M. Osada, S. Crossley, H. R. Lee, Y. Cui, Y. Hikita, and H. Y. Hwang, *Superconductivity in an Infinite-Layer Nickelate*, Nature **572**, 7771 (2019).

[6] S. Zeng et al., *Superconductivity in Infinite-Layer Nickelate $La_{1−x}Ca_xNiO_2$ Thin Films*, Sci. Adv. **8**, eabl9927 (2022).

[7] Q. Gu and H.-H. Wen, *Superconductivity in Nickel-Based 112 Systems*, The Innovation **3**, 100202 (2022).

[8] X. Wu, D. Di Sante, T. Schwemmer, W. Hanke, H. Y. Hwang, S. Raghu, and R. Thomale, *Robust $d_{x2-y2}$-Wave Superconductivity of Infinite-Layer Nickelates*, Phys. Rev. B **101**, 060504 (2020).

[9] H. Sun et al., *Signatures of Superconductivity near 80 K in a Nickelate under High Pressure*, Nature **621**, 7979 (2023).

[10] J. Hou et al., *Emergence of High-Temperature Superconducting Phase in the Pressurized $La_3Ni_2O_7$ Crystals*, Chin. Phys. Lett. (2023).

[11] Q. Qin and Y. Yang, *High-$T_c$ Superconductivity by Mobilizing Local Spin Singlets and Possible Route to Higher $T_c$ in Pressurized $La_3Ni_2O_7$*, Phys. Rev. B **108**, L140504 (2023).

[12] G. Wang et al., *Pressure-Induced Superconductivity in Polycrystalline $La_3Ni_2O_{7-δ}$*, arXiv:2309.173782v2 (2023).

[13] Z. Zhang, M. Greenblatt, and J. B. Goodenough, *Synthesis, Structure, and Properties of the Layered Perovskite $La_3Ni_2O_{7-δ}$*, J. Solid State Chem. **108**, 402 (1994).

[14] V. V. Poltavets, K. A. Lokshin, T. Egami, and M. Greenblatt, *The Oxygen Deficient Ruddlesden–Popper $La_3Ni_2O_{7−δ}$ (δ = 0.65) Phase: Structure and Properties*, Mater. Res. Bull. **41**, 955 (2006).

[15] V. V. Poltavets, K. A. Lokshin, S. Dikmen, M. Croft, T. Egami, and M. Greenblatt, *$La_3Ni_2O_6$: A New Double T′-Type Nickelate with Infinite $Ni^{1+/2+}O_2$ Layers*, J. Am. Chem. Soc. **128**, 9050 (2006).

[16] *Bjscistar*, http://www.bjscistar.com/page169?product_id=127.

[17] G. Shen et al., *Toward an International Practical Pressure Scale: A Proposal for an IPPS Ruby Gauge (IPPS-Ruby2020)*, High Press. Res. **40**, 299 (2020).

[18] T. Moriga, O. Usaka, I. Nakabayashi, T. Kinouchi, S. Kikkawa, and F. Kanamaru, *Characterization of Oxygen-Deficient Phases Appearing in Reduction of the Perovskite-Type $LaNiO_3$ to $La_2Ni_2O_5$*, Solid State Ion. **79**, 252 (1995).

[19] J. A. Alonso and M. J. Martínez-Lope, *Preparation and Crystal Structure of the Deficient Perovskite $LaNiO_{2.5}$, Solved from Neutron Powder Diffraction Data*, J. Chem. Soc. Dalton Trans. 2819 (1995).

[20] L. Jin, M. Lane, D. Zeng, F. K. K. Kirschner, F. Lang, P. Manuel, S. J. Blundell, J. E. McGrady, and M. A. Hayward, *LaSr$_3$NiRuO$_4$H$_4$: A 4d Transition-Metal Oxide–Hydride Containing Metal Hydride Sheets*, Angew. Chem. Int. Ed. **57**, 5025 (2018).
[21] L. Jin and M. A. Hayward, *Hole and Electron Doping of the 4d Transition-Metal Oxyhydride LaSr$_3$NiRuO$_4$H$_4$*, Angew. Chem. Int. Ed. **59**, 2076 (2020).
[22] Z. Xu, L. Jin, J.-K. Backhaus, F. Green, and M. A. Hayward, *Hole and Electron Doping of Topochemically Reduced Ni(I)/Ru(II) Insulating Ferromagnetic Oxides*, Inorg. Chem. **60**, 14904 (2021).
[23] S. Taniguchi, T. Nishikawa, Y. Yasui, Y. Kobayashi, J. Takeda, S. Shamoto, and M. Sato, *Transport, Magnetic and Thermal Properties of La$_3$Ni$_2$O$_{7-\delta}$*, J. Phys. Soc. Jpn. **64**, 1644 (1995).
[24] L. Li, X. Hu, Z. Liu, J. Yu, B. Cheng, S. Deng, L. He, K. Cao, D.-X. Yao, and M. Wang, *Structure and Magnetic Properties of the S = 3/2 Zigzag Spin Chain Antiferromagnet BaCoTe$_2$O$_7$*, Sci. China Phys. Mech. Astron. **64**, 287412 (2021).
[25] X. Xu, G. Cheng, D. Ni, X. Gui, W. Xie, N. Yao, and R. J. Cava, *Spin Disorder in a Stacking Polytype of a Layered Magnet*, Phys. Rev. Mater. **7**, 024407 (2023).
[26] V. Pardo and W. E. Pickett, *Metal-Insulator Transition in Layered Nickelates La$_3$Ni$_2$O$_{7-\delta}$ (δ = 0.0, 0.5, 1)*, Phys. Rev. B **83**, 245128 (2011).

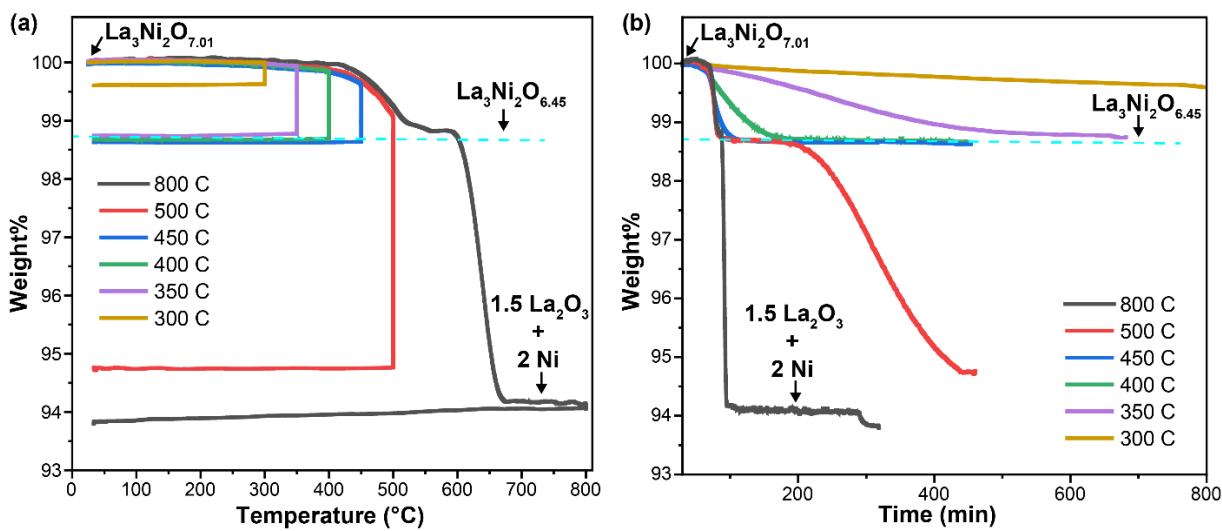

**Figure 1. The TGA Reduction Profile of La₃Ni₂O₇.** The La$_3$Ni$_2$O$_7$ samples were reduced in forming gas (5% H$_2$/Ar) under different reaction conditions. The resulting sample weight percents (obtained in the TGA) are plotted against a). temperature and b) time. (The time and temperature curves are from the same experiment.)

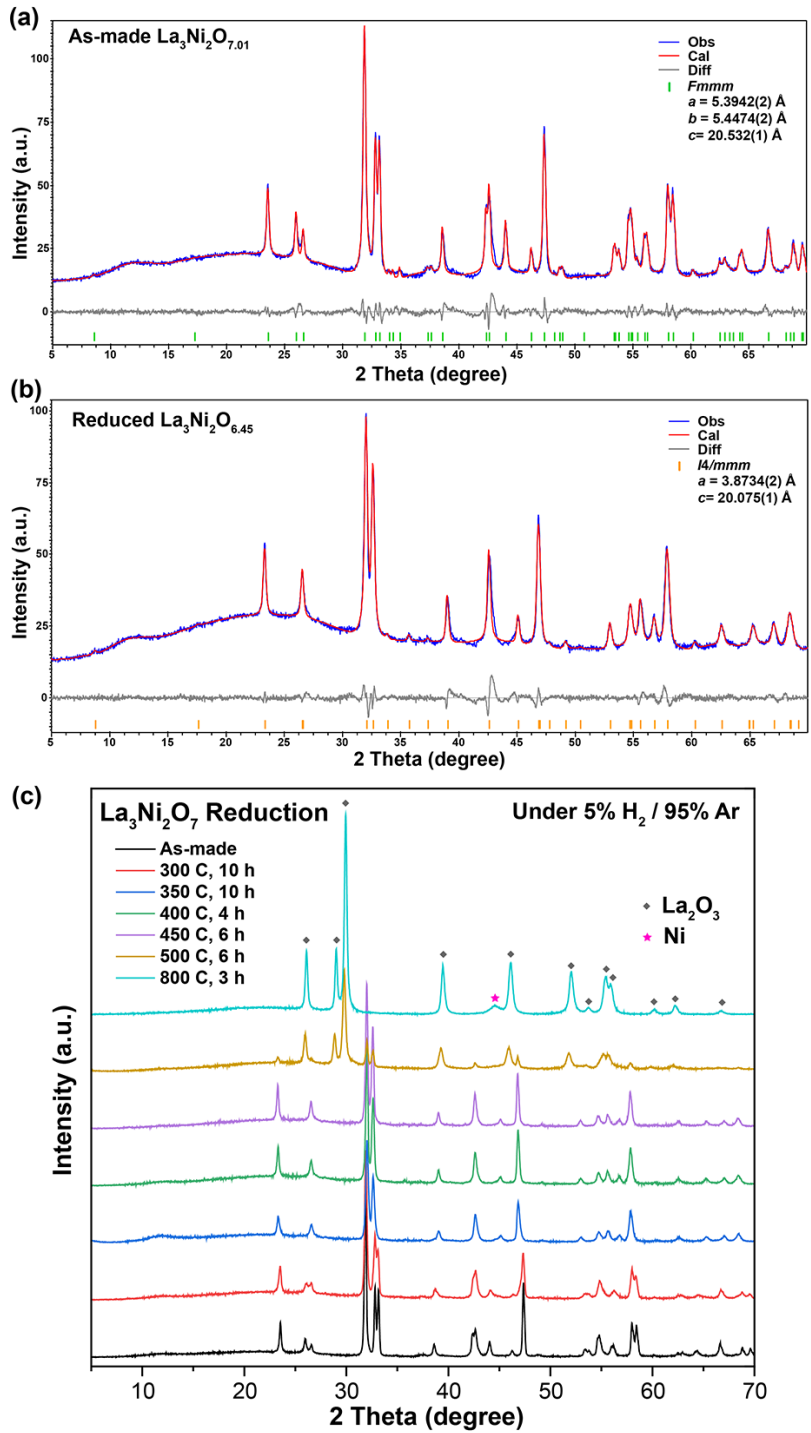

**Figure 2. The Structural Analysis.** The Le Bail fitting against the PXRD pattern collected from a) the as-made $La_3Ni_2O_{7.01}$ phase and b) the reduced $La_3Ni_2O_{6.45}$ phase; c) the PXRD patterns of the post-TGA samples made under different reduction conditions.

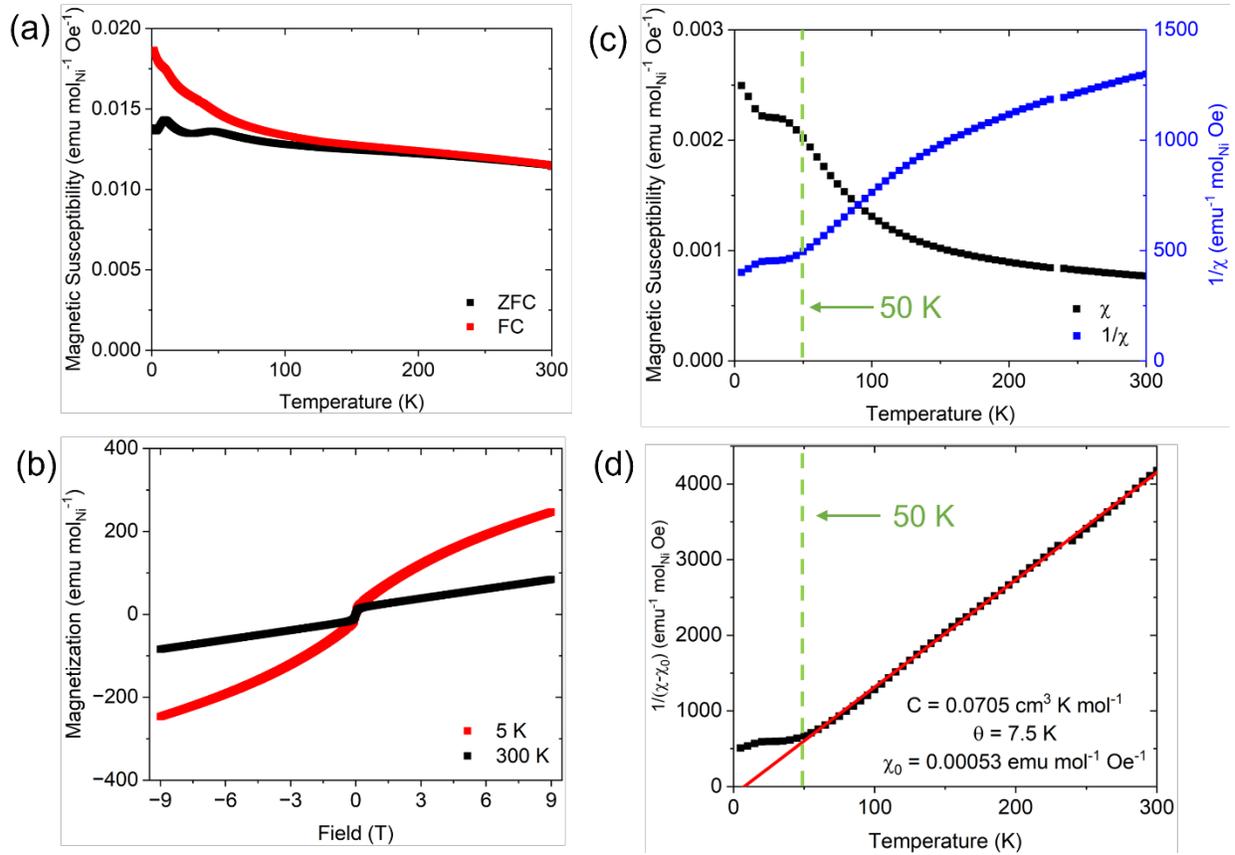

**Figure 3. The Magnetization of La$_3$Ni$_2$O$_{6.45}$.** a) Magnetic susceptibility derived from the temperature-dependent magnetization data plotted against temperature ($H$ = 1000 Oe); b) field-dependent magnetization data collected at 5 K and 300 K; Note that the 300 K data allows us to estimate that the amount ferromagnetic impurity (if elemental Ni is in the reduced material, it is a very small amount, 0.43% per mole of La$_{1.5}$NiO$_{3.225}$.) c) The ferro-subtracted magnetic susceptibility (black) and the inverse of magnetic susceptibility (blue); d) the Curie-Weiss fit to the high-temperature region of the $\chi_0$ corrected $1/\chi$ vs. temperature plot from panel c.

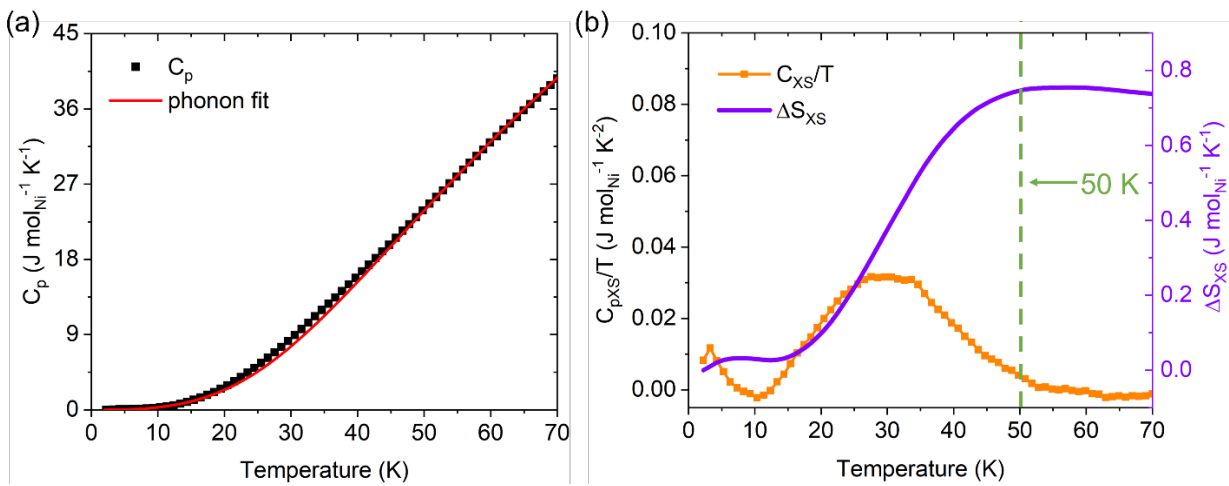

**Figure 4. The Heat Capacity of La$_3$Ni$_2$O$_{6.45}$.** a) The total heat capacity data C$_p$ with the high temperature region fitted by a modified two-component Debye equation to estimate the phonon contribution; b) The resulting C$_{mag}$/T (orange) data, as well as the excess entropy change ΔS$_{XS}$ (purple) plotted against temperature. The vertical dashed line at 50 K indicates that temperature.

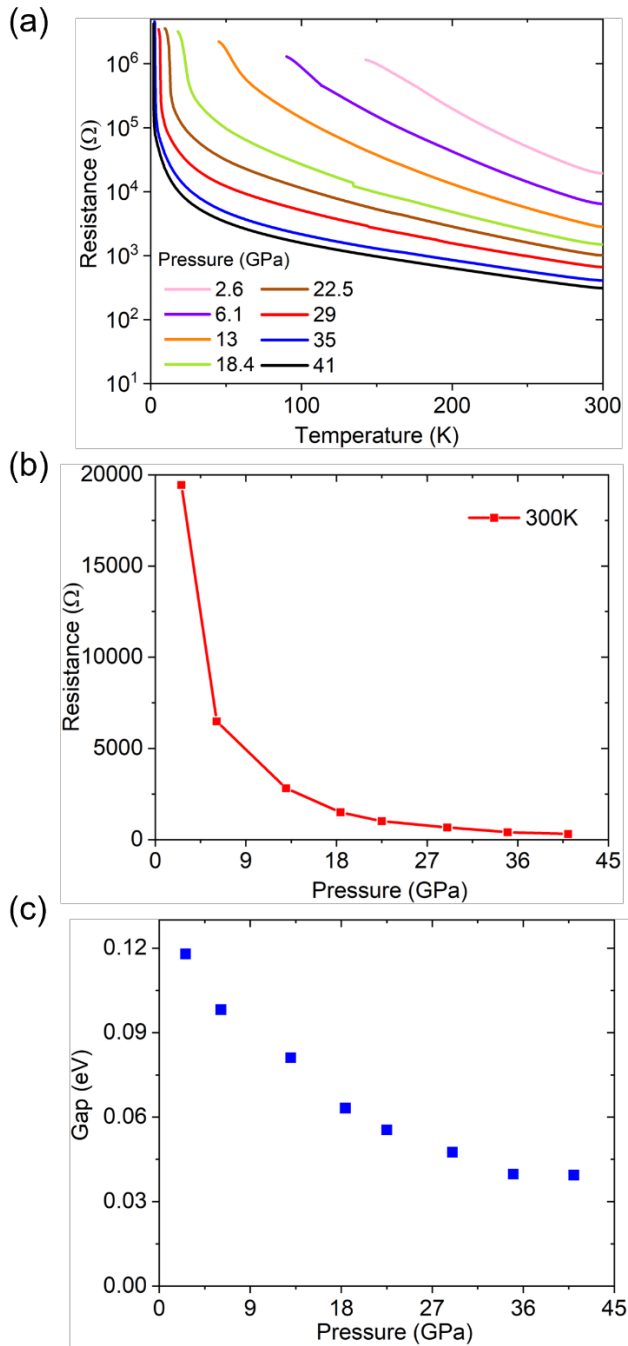

**Figure 5. Pressurized Resistance of La$_3$Ni$_2$O$_{6.45}$.** a) The temperature-dependent resistivity at various applied pressures; (b) The ambient temperature resistance plotted against applied pressure; (c) The resistively determined energy gap as a function of pressure.

# Supporting Information for

# Is La$_3$Ni$_2$O$_{6.5}$ a Bulk Superconducting Nickelate?

Ran Gao[1#], Lun Jin[1#], Shuyuan Huyan[2,3#], Danrui Ni[1*], Haozhe Wang[4], Xianghan Xu[1], Sergey L. Bud'ko[2,3], Paul Canfield[2,3], Weiwei Xie[4*] and Robert J. Cava[1*]

[1]Department of Chemistry, Princeton University, Princeton, New Jersey 08544, USA

[2]Ames National Laboratory, Iowa State University, Ames, IA 50011, USA

[3]Department of Physics and Astronomy, Iowa State University, Ames, IA 50011, USA

[4]Department of Chemistry, Michigan State University, East Lansing, Michigan 48824, USA

**\*** E-mails of corresponding authors: xieweiwe@msu.edu; danruin@princeton.edu; rcava@princeton.edu

# L.J., R.G. and S.H. contributed equally.

## Table of Contents

1. **Figure S1**. High-field fitting of magnetization (*M*) against field (*H*) data for ferro-subtraction. Note that all panels are plotted on the scale "per mole Ni", not "per mole formula unit".
2. **Figure S2**. Field dependent magnetization data of pure Ni powder at 300 K. Note that the data are plotted on the scale "per mole Ni", not "per mole formula unit".
3. **Figure S3.** Resistance of the reduced phase La$_3$Ni$_2$O$_{6.45}$ plotted against 1/T under different pressures.

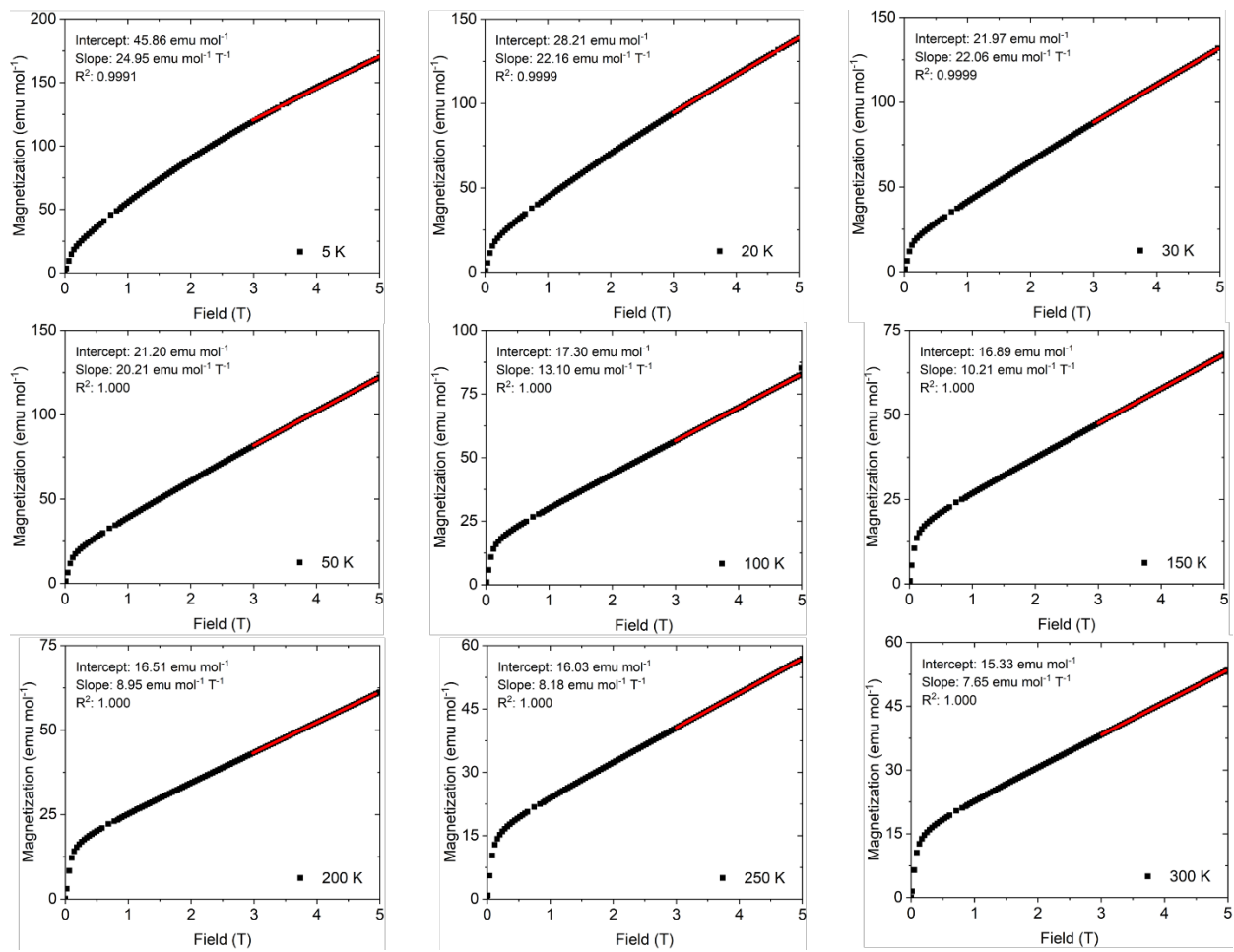

**Figure S1**. High-field fitting of magnetization (*M*) against field (*H*) data for ferro-subtraction. Note that all panels are plotted on the scale "per mole Ni", not "per mole formula unit".

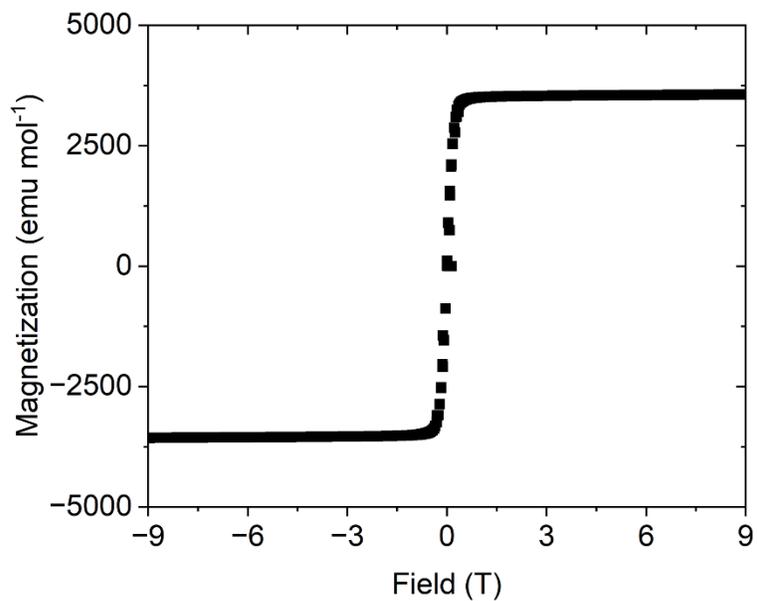

**Figure S2**. Field dependent magnetization data of pure Ni powder at 300 K. Note that the data are plotted on the scale "per mole Ni", not "per mole formula unit".

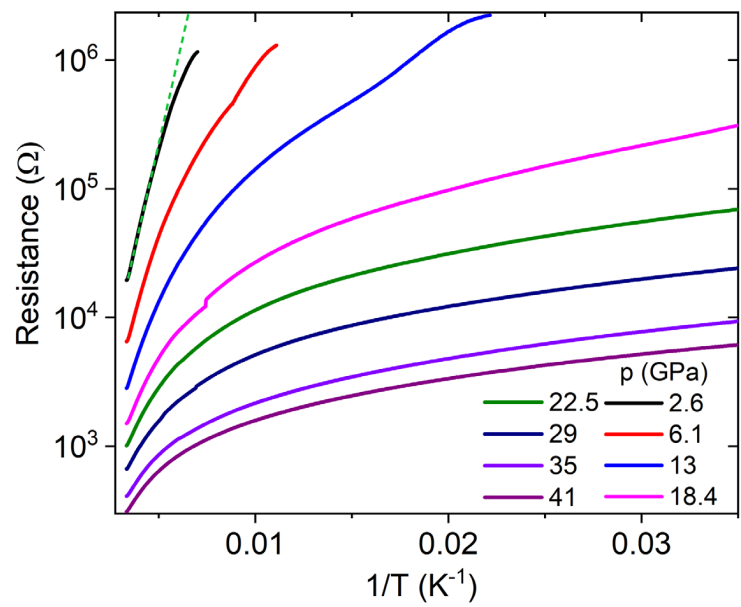

**Figure S3.** Resistance of the reduced phase La$_3$Ni$_2$O$_{6.45}$ plotted against 1/T under different pressures.